\documentclass[article]{agujournal2019}

\usepackage{amsmath}
\usepackage{graphicx}
\usepackage{psfrag}
\usepackage{stmaryrd}
\usepackage{amssymb}
\usepackage{wasysym}
\usepackage{float}
\usepackage{color}

\usepackage[T1]{fontenc}
\usepackage[french,english]{babel}
\usepackage{lmodern}
\usepackage{graphicx} 
\usepackage{times}
\usepackage{amsmath,amssymb}
\usepackage{color}

\newcommand{\cor}[1]{{#1}}

\newcommand{\la}{\left<}
\newcommand{\ra}{\right>}

\journalname{GRL}

\begin{document}

\title{Vertical structure of buoyancy transport by ocean baroclinic turbulence}


%

\authors{Julie Meunier\affil{1}, Benjamin Miquel\affil{2}, Basile Gallet\affil{1}}

\affiliation{1}{Universit\'e Paris-Saclay, CNRS, CEA, Service de Physique de l'Etat Condens\'e, 91191 Gif-sur-Yvette, France.}
\affiliation{2}{Univ Lyon, CNRS, Ecole Centrale de Lyon, INSA Lyon, Universit\'e Claude Bernard Lyon 1, LMFA, UMR5509, 69130, Ecully, France.}

\correspondingauthor{Basile Gallet}{basile.gallet@cea.fr}

\begin{keypoints}
\item We derive a prediction for the depth-dependence of the buoyancy flux associated with ocean baroclinic turbulence in an idealized setup.
\item The prediction is validated quantitatively by simulations of an idealized patch of ocean meant to resemble Southern Ocean conditions.
\item The prediction can be readily implemented into global models as a vertical profile for the Gent-McWilliams coefficient.
\end{keypoints}





\begin{abstract}
Ocean mesoscale eddies enhance meridional buoyancy transport, notably in the Antarctic Circumpolar Current where they contribute to setting the deep stratification of the neighboring ocean basins. The much-needed parameterization of this buoyancy transport in global climate models requires a theory for the overall flux, but also for its vertical structure inside the fluid column. Based on the quasi-geostrophic dynamics of an idealized patch of ocean hosting an arbitrary vertically sheared zonal flow, we provide a quantitative prediction for the vertical structure of the buoyancy flux without adjustable parameters. The prediction agrees quantitatively with meridional flux profiles obtained through numerical simulations of an idealized patch of ocean with realistic parameter values. This work empowers modelers with an explicit and physically based expression for the vertical profile of buoyancy transport by ocean baroclinic turbulence, as opposed to the common practice of using arbitrary prescriptions for the depth-dependence of the transport coefficients.
\end{abstract}

\section*{Plain Language Summary}
Ocean mesoscale vortices are turbulent structures tens of kilometers wide that play a central role in transporting tracers such as heat, salt and carbon. In the Southern Ocean, the associated buoyancy transport crucially sets the deep stratification of neighboring ocean basins. Because mesoscale vortices are not resolved by most state-of-the-art climate models, modelers resort to rather crude parameterizations where -- in the absence of a better theory -- the transport properties of the eddies are often assumed to be depth-invariant in the ocean interior. In this contribution we derive a quantitative and parameter-free prediction for the vertical structure of the turbulent buoyancy flux, which can be readily implemented in global models at little computational cost.

\section{Introduction}

The baroclinic instability of large-scale ocean currents generates mesoscale eddies that strongly enhance heat and tracer transport. In the Antarctic Circumpolar Current, the resulting turbulent buoyancy transport contributes to setting the slope of the Southern Ocean density surfaces and therefore the deep stratification of the neighboring ocean basins~\cite{Wolfe2010,Nikurashin11,Nikurashin12}. Mesoscale eddies have a core size comparable to the Rossby deformation radius, a length scale of the order of $60$~km at midlatitudes and $15$~km in the Southern Ocean, smaller than the coarse resolution of most global climate models. Parameterizing the transport induced by mesoscale eddies in such global models is thus crucial to obtain realistic ocean states that quantitatively reproduce the sloping density surfaces of the Southern ocean and the deep stratification of ocean basins. Physically-based parameterizations are inferred from the study of an isolated patch of ocean, where baroclinic turbulence has homogeneous statistics in the horizontal directions. The parameterization problem then consists in determining the scaling behavior of the overall diffusivity in terms of the various control parameters (shear flow magnitude, background stratification, bottom friction coefficient, etc.) but also the vertical structure of the various fluxes within the water column. Far more studies have addressed the former task~\cite{Phillips,Salmon,Salmon80,Larichev95,Held96,Arbic,Arbic2004b,Arbic2004a,Thompson06,Thompson07,Chang,Gallet2020,Gallet2021} than the latter~\cite{Stanley20,Zhang22,Yankovsky22}  in the ocean context. In the absence of a better theory many global models assume that the transport coefficients are depth-invariant in the ocean interior (see, e.g.~\citeA{Griffies05}), while other models consider surface-intensified coefficients with arbitrary prescriptions for their vertical structure (such as, e.g., assuming that the coefficients are proportional to the local squared buoyancy frequency~\cite{Ferreira05,Danabasoglu07,Gent11}). The latter assumption of surface-intensified  transport coefficients is at odds with idealized eddy-resolving channel simulations, which point to a bottom-enhanced buoyancy transport coefficient instead~\cite{Abernathey13} (the so-called Gent-McWilliams coefficient, see below). 

\begin{figure}
    \centerline{\includegraphics[width=10 cm]{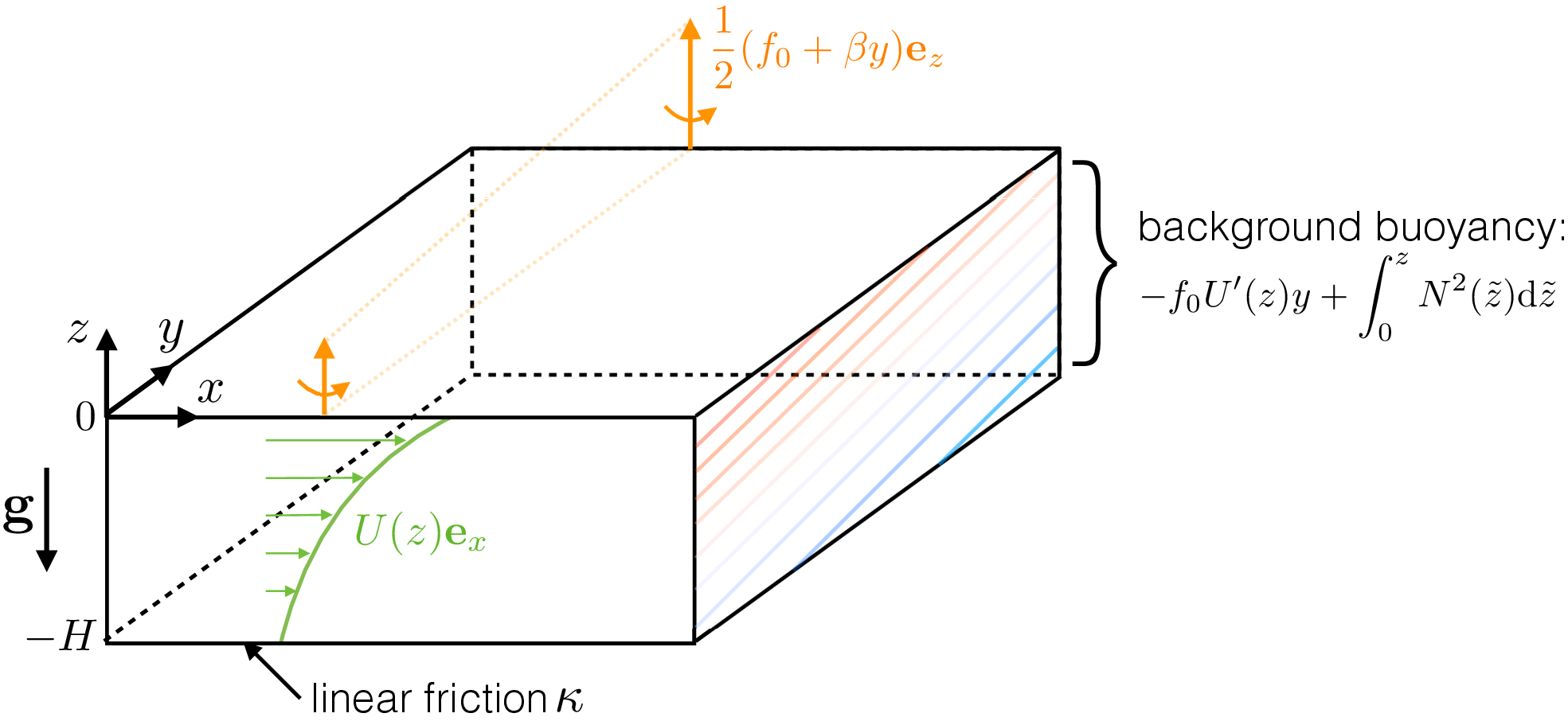} }
   \caption{\textbf{An idealized patch of ocean.} A layer of fluid is subject to global rotation at a rate that varies linearly with the meridional coordinate $y$. The fluid is density stratified with an arbitrary profile $N(z)$ for the buoyancy frequency. The background zonal shear flow has an arbitrary profile $U(z)$. This flow coexists with a background meridional buoyancy gradient. Friction damps kinetic energy on the ocean floor. \label{fig:schematic}}
\end{figure}

To improve upon this unsatisfactory state of the art, in this Letter we derive a parameter-free prediction for the vertical structure of the turbulent buoyancy flux within the water column.
We consider an idealized patch of ocean with arbitrary background zonal shear flow and stratification, and $\beta \neq 0$, see Figure~\ref{fig:schematic}. Water occupies a volume $(x,y,z)\in [0,L]^2 \times [-H,0]$ with a stress-free boundary at $z=0$ and a linear-friction boundary condition at $z=-H$, in a frame rotating around the vertical axis with a local Coriolis parameter $f_0+\beta y$, where $y$ denotes the meridional (North-South) coordinate. The fluid layer is density-stratified with an arbitrary profile $N(z)$ for the buoyancy frequency, and we restrict attention to a single stratifying agent. We focus on the quasi-geostrophic (QG) regime arising for fast rotation and strong stratification~\cite{Venaille11,Salmonbook,Vallisbook}. The base flow consists of an arbitrary zonal velocity profile $U(z)$ in thermal wind balance with a $z$-dependent meridional buoyancy gradient $\partial_y B=-f_0 U’(z)$, where the prime symbol denotes a vertical derivative. We consider arbitrary departures from this base state with periodic boundary conditions in the horizontal directions. We denote as $p(x,y,z,t)$ the departure from the base pressure field, with $u=-p_y$ the departure zonal velocity, $v=p_x$ the departure meridional velocity, $b=f_0 \, p_z$ the departure buoyancy and $w$ the subdominant (ageostrophic) vertical velocity. Non-dimensionalizing time and space using $|f_0|^{-1}$ and $H$, the dimensionless base flow is written as $U/|f_0|H=Ro \, {\cal U}(z)$, where $Ro=|U(0)/f_0 H|$ is the Rossby number associated with the surface speed of the base flow and ${\cal U}(z)$ denotes the base-flow profile normalized at the surface ($|{\cal U}(0)|=1$). For brevity we use the same symbols for the dimensionless variables. 

Consider a tracer $\tau$ stirred by the 3D flow and subject to horizontally uniform gradients (at lowest order in $Ro$) $G_y^{(\tau)}(z)$ and $G_z^{(\tau)}(z)$ in the meridional and vertical directions, respectively. The QG evolution equation for $\tau$ reads:
\begin{equation}
\partial_t \tau + Ro \, {\cal U}(z) \, \tau_x + J(p,\tau) =  - p_x G_y^{(\tau)}(z) - w G_z^{(\tau)}(z)  + {\cal D}_\tau \, , \label{eq:constau}
\end{equation}
where the Jacobian is $J(g,h)=g_x h_y - g_y h_x$ and ${\cal D}_\tau$ denotes small-scale diffusion.

Denoting with an overbar $\overline{\cdot}$ a time average together with a horizontal area average, the eddy-induced meridional and vertical fluxes of $\tau$ are related to the background gradients by a Gent-McWilliams/Redi (GM/R) diffusion tensor~\cite{Redi82,Gent90,Griffies98,Mcdougall2001,Gent11}:
\begin{eqnarray}
\left( \begin{matrix}
\overline{v\tau} \\
\overline{w\tau}
\end{matrix}  \right) =
\left[\begin{matrix}
-K_R &  (K_{GM}-K_R) {\cal S}\\
-   (K_{GM}+K_R){\cal S}& - K_R {\cal S}^2\\
\end{matrix} \right]
\left( \begin{matrix}
G_y^{(\tau)} \\
G_z^{(\tau)}
\end{matrix}  \right)  \,  \quad \label{GMRedi}
\end{eqnarray}
where the Redi diffusivity $K_R(z)$ encodes diffusion along the mean isopycnal direction, the GM coefficient $K_{GM}(z)$ encodes the advective (or skew-diffusive) transport, and we denote the isopycnal slope of the base state as ${\cal S}(z)=Ro\, {\cal U}'/N^2$. While~(\ref{GMRedi}) is often introduced based on physical intuition and educated guesses, we have recently proposed a direct derivation of this diffusion tensor from the quasi-geostrophic dynamics of the present system~\cite{Meunier23}. For completeness we briefly recall a few results from this recent study.

The quasi-geostrophic potential vorticity (QGPV) $q=\Delta_\perp p + \partial_z \left[ {p_z}/{N^2(z)} \right]$ is governed by equation (\ref{eq:constau}) with $\tau=q$, $G_z^{(q)}=0$ and $G_y^{(q)}=\tilde{\beta}-{\cal S}'(z)$, while buoyancy is governed by (\ref{eq:constau}) with $\tau=b$, $G_z^{(b)}=N^2$ and $G_y^{(b)}=-Ro \, {\cal U}'$. Substitution of these background gradients into the flux-gradient relation (\ref{GMRedi}) indicates that $K_{GM}(z)$ and $K_R(z)$ can alternatively be thought of as the effective diffusivities associated with the meridional transport of $b$ and $q$, respectively:
\begin{equation}
K_{GM}=-\frac{\overline{vb}}{G_y^{(b)}}=\frac{\overline{vb}}{Ro \,  \, {\cal U}'} \, , \qquad K_{R}=-\frac{\overline{vq}}{G_y^{(q)}}=\frac{\overline{vq}}{{\cal S}'(z)-\tilde{\beta}} \, . \label{eq:altdefK}
\end{equation}

Because the vertical velocity vanishes at the surface, the governing equations for $q$ and $b$ admit the same limiting form as one approaches the top boundary. Both tracers are advected by the surface horizontal flow, fluctuations being induced by distortions of a horizontally homogeneous background meridional gradient. The associated meridional diffusivity is thus equal for $b$ and $q$ at the surface: $\overline{vb}(0)/G_y^{(b)}(0)=\overline{vq}(0)/G_y^{(q)}(0)$. Provided the friction coefficient is small, the same holds near the bottom boundary, at a depth $z=-1^+$ located just above the bottom Ekman layer: $\overline{vb}(-1^+)/G_y^{(b)}(-1^+) \simeq \overline{vq}(-1^+)/G_y^{(q)}(-1^+)$ (while the friction-induced vertical pumping velocity is crucial for damping kinetic energy through the stretching of planetary vorticity, it has a negligible direct contribution to buoyancy transport for low drag coefficient). Using (\ref{eq:altdefK}) we recast these equalities as:
\begin{equation}
K_{GM}(0)  =  K_R(0) \, ,    \qquad K_{GM}(-1^+) \simeq K_R(-1^+) \, . \label{eq:eqK}
\end{equation}
The two equalities in (\ref{eq:eqK}) are illustrated numerically in~\citeA{Meunier23}. An additional constraint on $K_{GM}$ and $K_R$ is obtained by substituting the definition of $q$ into the meridional QGPV flux $\overline{vq}$. After a few integration by parts one obtains the Taylor-Bretherton relation $\overline{vq}=\mathrm{d}(\overline{vb}/N^2)/\mathrm{d}z$~\cite{Taylor15,Bretherton66,Smith09,Dritschel08,Young12}, and expressing the meridional fluxes using (\ref{eq:altdefK}):
\begin{equation}
K_R({\cal S}'-\tilde{\beta})=\frac{\mathrm{d}}{\mathrm{d} z} (K_{GM} \, {\cal S}) \, . \label{TBrelation}
\end{equation}
In the following we show that the constraints (\ref{eq:eqK}-\ref{TBrelation}) allow for a perturbative derivation of the vertical structure of the eddy-induced buoyancy flux within the water column in two situations of interest.

\section{Case I: The impact of weak $\beta$ on Eady turbulence} 

\begin{figure}
      \centerline{\includegraphics[width=18 cm]{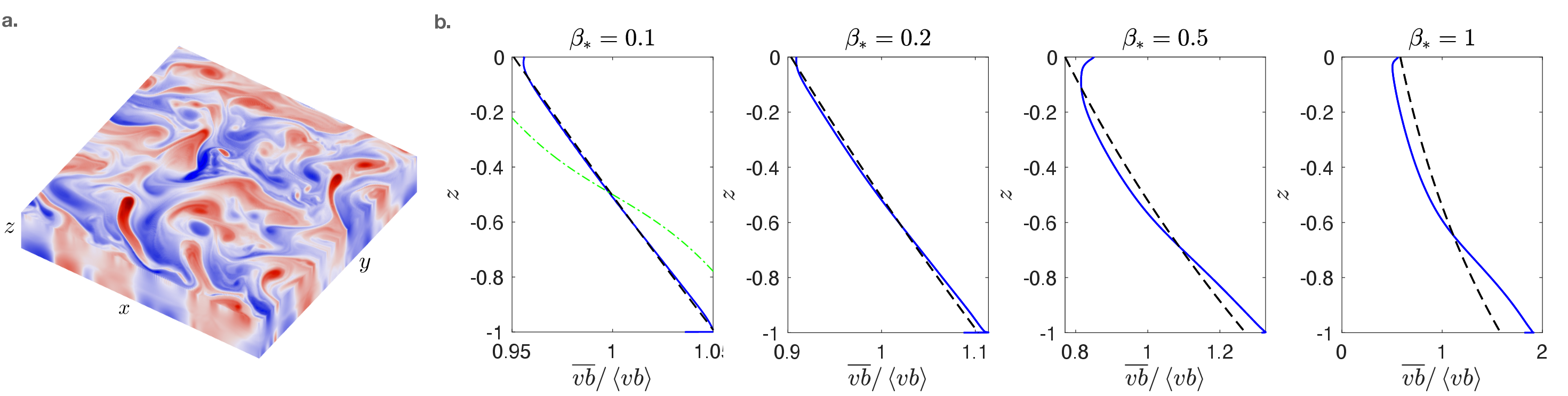} }
   \caption{\textbf{An illustrative example: the Charney model with weak $\beta$. a.} Snapshot of the departure buoyancy field $b$ for $\beta_*=0.5$ (large values in red, low values in blue). \textbf{b.} Vertical structure of the meridional buoyancy flux in the equilibrated state for increasing $\beta_*$ (solid line: DNS, dashed line: perturbative prediction (\ref{vbexpprediction})). The agreement with the prediction is excellent in the perturbative regime $\beta_*\ll 1$ and deteriorates somewhat as $\beta_*$ reaches ${\cal O}(1)$ values. The perturbative prediction performs always better than the common practice of parameterizing turbulent transport using a depth-invariant $K_{GM}$, which corresponds to depth-invariant $\overline{vb}$ for the present setup. The prediction (\ref{vbexpprediction}) also performs better than using the meridional flux associated with the most unstable eigenmode, computed perturbatively for weak $\beta_*$ and represented for $\beta_*=0.1$ as a green dash-dotted line.\label{fig:Charney}}
\end{figure}

The QG Eady model corresponds to depth-independent stratification $N^2$ and shear ${\cal U}'$ (linear zonal velocity profile, ${\cal U}(z) =z+1$), together with $\beta=0$. As discussed in \citeA{Gallet2022}, there is no background PV gradient in this setup and therefore a solution can be obtained by assuming $q=0$ in the bulk of the domain. The meridional QGPV flux then vanishes, and from relation (\ref{TBrelation}) we conclude that the meridional buoyancy flux, and thus $K_{GM}$, are independent of $z$. 

As established in~\citeA{Meunier23}, $K_R(z)$ is given by the Taylor-Kubo eddy diffusivity coefficient associated with the horizontal geostrophic flow. That is, at every depth $z$ the coefficient $K_R(z)$ is given by the integral of the Lagrangian correlation function of the horizontal geostrophic flow. Because in the low-drag limit the Eady flow barotropizes, we expect the horizontal geostrophic flow to be depth-invariant, which leads to $K_R$ being independent of $z$.
Using the boundary relation (\ref{eq:eqK}) we conclude that the GM and Redi coefficients are depth-invariant and equal to one another. The low-drag Eady model thus represents one limiting situation for which the depth-invariance and equality of the GM and Redi coefficients can be established. \cor{We stress the fact that the equality of the GM and Redi coefficients has been established based on the properties of the low-drag equilibrated state, namely barotropization, and the theory presented below is really a theory for such a low-drag equilibrated -- or 'turbulent' -- state. By contrast, the theory would not hold to predict the vertical structure of an eigenmode obtained using linear stability analysis, whose transport properties typically display strong depth-dependence (see Supplementary Information).}

Consider now the impact of a weak planetary vorticity gradient on Eady turbulence, that is, a Charney model with weak $\beta$. Within QG, the impact of $\beta$ is characterized by the product of $\beta$ with the squared deformation radius over the typical velocity of the background shear flow~\cite{Charney47,Thompson07,Gallet2021,Chang21}. We thus define the ($z$-invariant) parameter $\beta_*=\tilde{\beta} N^2/Ro$. In the perturbative regime $\beta_* \ll 1$ the correction to the $z$-invariant $\beta_*=0$ situation is small, and a standard expansion leads to
$K_{R}(z)=K_{GM}(z) [1+{\cal O}(\beta_*)]$, where the ${\cal O}(\beta_*)$ correction vanishes both at the top and at the bottom boundary in the low-drag weakly diffusive regime, see equation (\ref{eq:eqK}). Substitution into (\ref{TBrelation}) yields $K_{GM}'(z) = - \beta_* K_{GM}(z) + {\cal O}(\beta_*^2)$ and, neglecting the ${\cal O}(\beta_*^2)$ correction, $K_{GM}(z) = \text{const.} \times  \, e^{-\beta_* z}$. Substituting this expression for $K_{GM}(z)$ into (\ref{eq:altdefK}) and denoting the overall buoyancy flux as $\left< vb \right> = \int_{-1}^0 \overline{vb}(\tilde{z})\mathrm{d}\tilde{z}$, we obtain a parameter-free prediction for the vertical structure $\overline{vb}(z)/\la vb \ra$ of the meridional buoyancy flux:
\begin{equation}
\frac{\overline{vb}(z)}{\la vb \ra} = \frac{\beta_*}{e^{\beta_*}-1} \, e^{-\beta_* z} \, . \label{vbexpprediction}
\end{equation}

To test this perturbative prediction, we have performed numerical simulations of this setup in the QG regime with periodic boundary conditions in the horizontal directions.
As detailed in the Supporting Information (see also \citeA{Meunier23}), our numerical approach consists in time-stepping a set of primitive-like equations with tailored $\beta$ terms that are compatible with the horizontal periodic boundary conditions. Importantly, these tailored terms reduce to the standard $\beta$ terms in the QG limit. Because we focus on parameter values that are strongly QG, this approach is equivalent to (but more convenient than) directly solving the QG system. In Figure~\ref{fig:Charney} we plot the vertical structure of the meridional buoyancy flux, $\overline{vb}(z)/\la vb \ra$, for increasing $\beta_*$. The numerical profiles are in excellent agreement with the parameter-free prediction (\ref{vbexpprediction}) for low $\beta_*$. As expected, the perturbative prediction deteriorates somewhat as $\beta_*$ increases up to $\beta_*=1$. In the next section we show that the perturbative regime accurately captures the typical oceanic situation, characterized by surface-intensified baroclinic turbulence.

\begin{figure}
    \centerline{\includegraphics[width=10 cm]{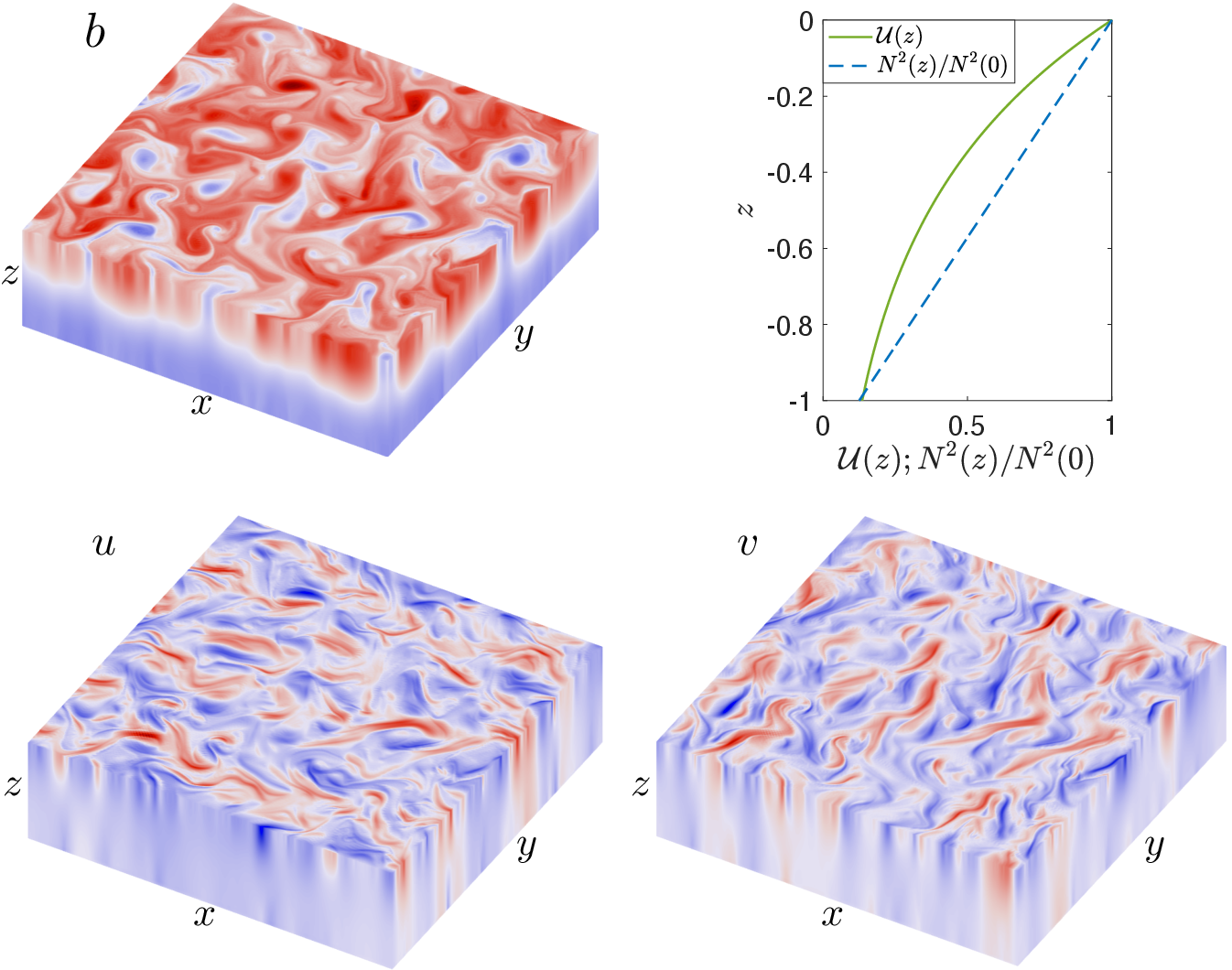} }
   \caption{\textbf{Surface-intensified baroclinic turbulence} from the base run, meant to ressemble a patch of the Antarctic Circumpolar Current (positive values in red and negative values in blue, background profiles in the upper-right panel).   \label{fig:snapshots_fneg}}
\end{figure}

\section{Case II: Surface-intensified shear and stratification} 

The perturbative approach developed in the preceding section is based on the small value of the background meridional PV gradient: when low-drag baroclinic turbulence is subjected to a weak meridional PV gradient, the profile of $K_{GM}(z)$ can be inferred by inserting $K_R(z) \simeq K_{GM}(z)$ into the Taylor-Bretherton relation (\ref{TBrelation}).

The meridional PV gradient associated with $\beta$ is modest in a typical oceanic setting, which may suggest that one can again use the approximate relation $K_R(z) \simeq K_{GM}(z)$ in the bulk of the domain. However, the PV gradient associated with the $z$-dependent shear profile is much greater (see figure 6 of \citeA{Smith09}). Fortunately, ocean baroclinic turbulence is surface-intensified and the largest shear-induced meridional PV gradient arises in the upper region of the fluid column, where the approximate equality $K_R \simeq K_{GM}$ holds by virtue of the near-surface relation (\ref{eq:eqK}). Once again, one can thus substitute the approximate relation $K_R(z) \simeq K_{GM}(z)$ into the Taylor-Bretherton relation (\ref{TBrelation}) to compute the profile of $K_{GM}$, this time perturbatively in distance from the upper boundary. We conclude that a useful approximation to the vertical dependence of $K_{GM}$ should be obtained by substituting $K_R(z) \simeq K_{GM}(z)$ into (\ref{TBrelation}) throughout the entire water column. \cor{As for Case I, one way to derail this procedure would be to have a surprisingly large meridional PV flux arise in the interior of the domain despite the very weak PV gradient. This typically happens for a nearly marginal eigenmode, but not for the present low-drag equilibrated -- or `turbulent' -- states. In that respect the theory below is really a theory for such equilibrated baroclinic turbulence.}

\cor{After re-arranging, the substitution of $K_R(z) \simeq K_{GM}(z)$ into (\ref{TBrelation})} leads to the following ODE for the vertical structure of the GM coefficient:
\begin{equation}
\frac{\mathrm{d}}{\mathrm{d} z}  \ln K_{GM} = - \frac{\tilde{\beta}}{{\cal S}(z)}\, . \label{ODEKGM}
\end{equation}
This relation points to the crucial role of $\beta$ in setting the vertical structure of the eddy-induced buoyancy flux: according to (\ref{ODEKGM}) the common assumption of a depth-invariant GM coefficient is valid for $\beta=0$ only. For arbitrary $\beta$ equation (\ref{ODEKGM}) can be integrated into:
\begin{equation}
K_{GM}(z) = \text{const.} \times \exp \left[- \int_0^z \frac{\tilde{\beta}}{{\cal S}(\tilde{z})} \mathrm{d}\tilde{z} \right]\, . \label{KGMgeneral}
\end{equation}
We have obtained an explicit expression for the vertical structure of the GM coefficient in terms of the vertical profiles of background stratification and shear. Using equation (\ref{eq:altdefK}), the expression (\ref{KGMgeneral}) can be recast into a parameter-free prediction for the vertical structure of the meridional buoyancy flux:
\begin{eqnarray}
\frac{\overline{vb}(z)}{\la vb \ra} & = &  \frac{{\cal U}'(z) \exp \left[- \int_0^z \frac{\tilde{\beta}}{{\cal S}(\tilde{z})} \mathrm{d}\tilde{z} \right]}{\int_{-1}^0 {\cal U}'(\tilde{z}) \exp \left[- \int_0^z \frac{\tilde{\beta}}{{\cal S}(\tilde{z})} \mathrm{d}\tilde{z} \right]  \mathrm{d}z}\, . \label{vbgeneral} 
\end{eqnarray}

To test the prediction (\ref{vbgeneral}) we have performed numerical simulations of surface-intensified baroclinic turbulence with parameter values typical of the Antarctic Circumpolar Current (ACC). The dimensionless stratification profile is linear in $z$ and surface intensified,  $N^2(z)=a_0+a_1 \, z$, with constant coefficients $a_0$ and $a_1$. The shear flow has an exponential profile ${\cal U}(z)=s\, e^{z/\ell}$ with an e-folding scale $\ell$ (in units of $H$) and a sign prefactor $s=+1$ for an eastward flow and $s=-1$ for a westward one. We perform a base run with dimensionless QG parameter values similar to the situation addressed by \citeA{Smith09}: dimensional magnitude $|f_0|=1.23 \times 10^{-4}$~s$^{-1}$ for the Coriolis parameter and $\beta=1.23 \times 10^{-11}$~m$^{-1}$.s$^{-1}$ for the planetary vorticity gradient (corresponding to a latitude of $57.5^o$S), depth of fluid equal to $H=4000$~m, eastward shear flow with surface speed $U(0)=0.15$~m.s$^{-1}$ and e-folding scale of $2000$~m. The dimensional buoyancy frequency ranges from $8.7 \times 10^{-4}$~s$^{-1}$ at the bottom to $2.46 \times 10^{-3}$~s$^{-1}$ at the surface, which corresponds to a Rossby deformation radius $\lambda \simeq 19$~km based on the rough WKB estimate $\lambda/H=\int_{-1}^0 N(z) \mathrm{d}z /\pi$ (recalling that $N(z)$ is non-dimensionalized with $|f_0|$). In terms of dimensionless parameters, these values translate into a Rossby number $Ro=0.3$, a vertical scale $\ell=0.5$ for the shear flow, a dimensionless planetary vorticity gradient $\tilde{\beta}=4.0 \times 10^{-4}$ and stratification coefficients $a_0=400$ and $a_1=350$. To ensure that the base numerical run indeed corresponds to the fully QG regime, we have used values for $Ro$ and $\tilde{\beta}$ that are smaller by a factor of $10$ (that is, we use $Ro=0.03$ and $\tilde{\beta}=4.0 \times 10^{-5}$), which leaves invariant the dissipation-free QG dynamics.

Together with this base run we have performed a run without $\beta$ and a run with $\beta > 0$ and a westward base flow. These additional runs are performed with slightly larger stratification ($a_0=800$ and $a_1=700$) using the inferred values $Ro=0.3$ and $\tilde{\beta}=4.0 \times 10^{-4}$ for the dimensionless buoyancy and planetary vorticity gradients. Finally, we have repeated similar runs using the larger value $\ell=1$ for the e-folding scale of the shear (see Supporting Information for the values of the other parameters).

\begin{figure}
    \centerline{\includegraphics[width=10 cm]{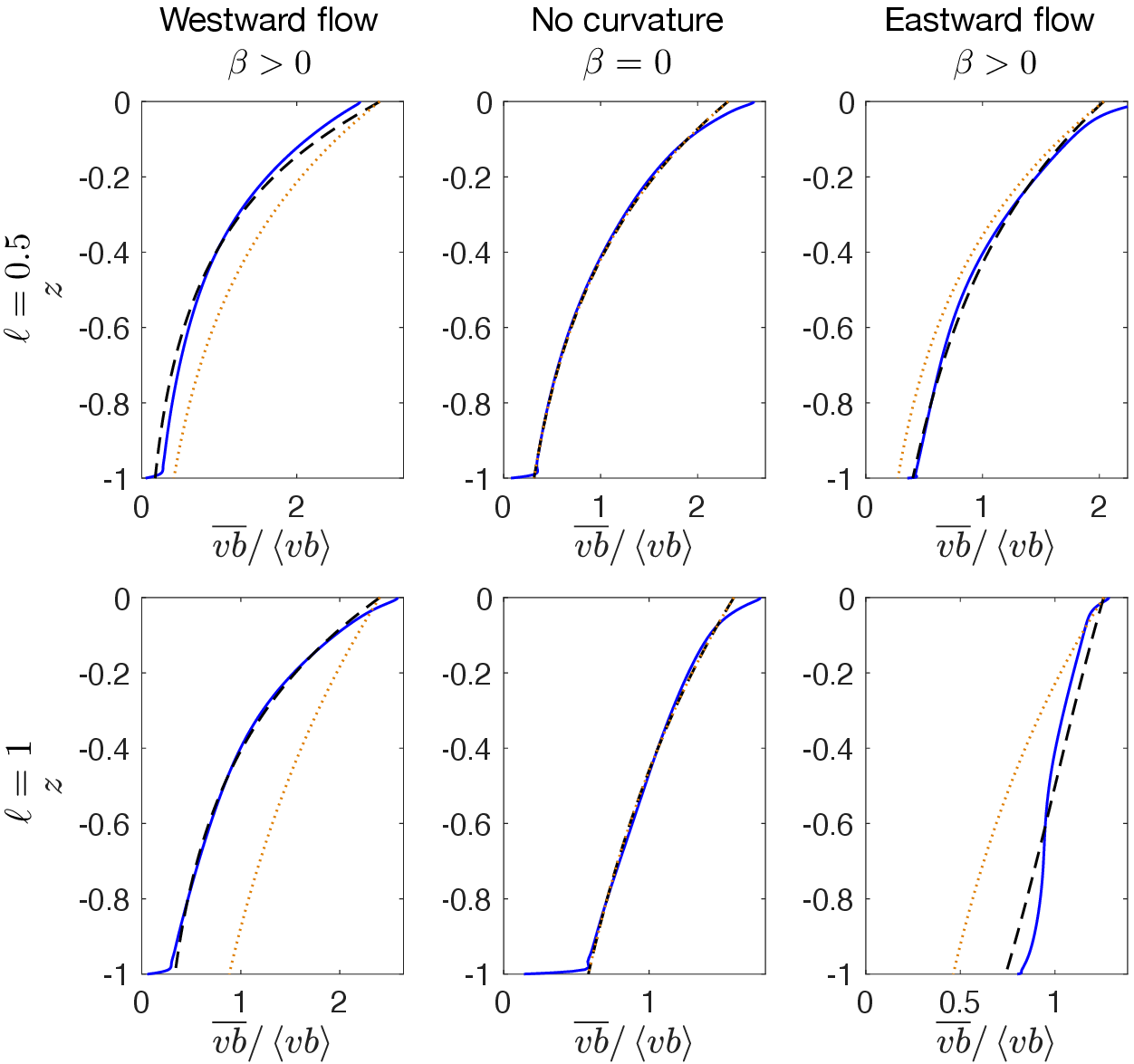} }
   \caption{\textbf{Vertical structure of the meridional buoyancy flux} for an eastward flow, for a westward flow and for the case $\beta=0$, using either $\ell=0.5$ or $\ell=1$. The solid line is the profile extracted from the numerical runs. The dashed line is the theoretical prediction (\ref{vbparticular}). The orange dotted-line corresponds to the uniform-$K_{GM}$ model that matches the surface value of the full prediction (\ref{vbparticular}) (see text for details). The prediction (\ref{vbparticular}) reduces to a depth-invariant $K_{GM}$ when $\beta=0$, which agrees accurately with the $\beta=0$ numerical profiles. For $\beta \neq 0$ the prediction (\ref{vbparticular}) departs from a uniform $K_{GM}$ and agrees well with the numerical profiles. \label{fig:vbprofiles}}
\end{figure}

In Figure~\ref{fig:snapshots_fneg} we provide snapshots of the buoyancy and velocity fields in the equilibrated state of the base run. As expected the turbulence is surface intensified and so is the meridional buoyancy flux $\overline{vb}$, provided in the upper-right panel of Figure~\ref{fig:vbprofiles}. Substituting the linear profile for $N^2(z)$ and the exponential profile for ${\cal U}(z)$ into expression (\ref{vbgeneral}), the theoretical prediction for the vertical structure of the meridional buoyancy flux becomes:
\begin{equation}
\frac{\overline{vb}(z)}{\la vb \ra} = \frac{\exp \left\{ \frac{z}{\ell} + \frac{s\, \tilde{\beta} \ell^2}{Ro} [a_0+a_1(z+\ell)] e^{-\frac{z}{\ell}} \right\}}{\int_{-1}^0     \exp \left\{ \frac{z}{\ell} + \frac{s\,  \tilde{\beta} \ell^2}{Ro} [a_0+a_1(z+\ell)] e^{-\frac{z}{\ell}} \right\}  \mathrm{d}z} \, . \label{vbparticular}
\end{equation}
We compare this prediction to the numerically determined meridional flux profiles in Figure~\ref{fig:vbprofiles}. The agreement is very good for both values of $\ell$, both with and without $\beta$, and for both eastward and westward flows. For $\beta=0$ the theoretical prediction is that of a depth-invariant GM coefficient, and thus a meridional buoyancy flux that inherits the vertical structure ${\cal U}'(z)$ of the background shear. The good agreement with the numerical profiles validates this prediction and indicates that a depth-invariant GM coefficient is indeed an excellent parameterization when $\beta=0$. For $\beta \neq 0$, however, the prediction (\ref{vbparticular}) departs from the common practice of using a depth-invariant GM coefficient. In all cases the prediction (\ref{vbparticular}) better captures the vertical structure of $\overline{vb}$, without adjustable parameters (see Figure~\ref{fig:vbprofiles}). The difference between the two predictions -- equation (\ref{vbparticular}) versus uniform $K_{GM}$ -- is modest for $\ell=0.5$ and greater for $\ell=1.0$. In particular, using a uniform $K_{GM}$ would lead to the same vertical structure for the buoyancy flux regardless of whether the base flow is directed eastward or westward. By contrast, the numerical data indicate that the vertical structure strongly depends on the direction of the base flow: for $\ell=0.5$ the bottom-to-top meridional flux ratio, evaluated as $\overline{vb}(-0.95)/\overline{vb}(-0.05)$, is 12\% for a westward flow and 24\% for an eastward flow. For $\ell=1$ this ratio is 15\% for a westward flow and 73\% for an eastward flow. 
For a visual illustration of these differences, we represent in Figure~\ref{fig:vbprofiles} the uniform-$K_{GM}$ prediction as an orange dotted-line for comparison with the present prediction, demanding that the two predictions be equal at the top surface $z=0$.

\section{Conclusion} 
The predictions (\ref{KGMgeneral}) and (\ref{vbgeneral}) are based on a perturbative approach that holds in the near-surface and near-bottom regions of the fluid column for arbitrary meridional potential vorticity gradient, and throughout the entire fluid column when the meridional potential vorticity gradient is weak. The present perturbative framework is useful for baroclinic turbulence in the ocean, where the shear flow and meridional PV gradient are boundary-intensified, with weaker PV flux in the interior.
One can then combine the Taylor-Bretherton relation between the buoyancy and PV fluxes with the near-equality of the GM and Redi coefficients in the vicinity of the boundaries. This leads to a prediction for the vertical structure of the buoyancy flux that agrees well with the profiles extracted from direct numerical simulations, see Fig.~\ref{fig:vbprofiles}. It would be interesting to further investigate the range of validity of the predictions (\ref{KGMgeneral}) and (\ref{vbgeneral}) beyond the present oceanographically relevant situations. For instance, a system with a vanishing meridional buoyancy gradient at the bottom $(G_y^{(b)}(-1)=0)$ may emphasize the role of bottom friction and disrupt the relation $K_{GM}(-1^+) \simeq K_R(-1^+)$. More generally, while surprisingly successful the present perturbative approach should probably be used with caution whenever the exponential factor in (\ref{KGMgeneral}) varies by much more than a factor of two within the water column.

For eastward shear flows (positive shear) the right-hand side of equation~(\ref{ODEKGM}) is negative: $K_{GM}(z)$ is greater at depth according to both the theory and the numerics, even though the turbulence is surface-intensified. This prediction is fully compatible with the $K_{GM}$-profile reported by~\citeA{Abernathey13} and challenges models where the $K_{GM}$-profile is assumed to be proportional to the profile of $N^2(z)$~\cite{Ferreira05}.
The present results also seem to invalidate the idea that the vertical structure of the flux could be governed by a single baroclinic mode~\cite{Stanley20}. Indeed, the modal decomposition~\cite{Flierl} is the same for all panels of Fig.~\ref{fig:vbprofiles} and yet the buoyancy flux profiles differ strongly between panels.

Another idea put forward in the atmospheric context is that the flux profiles in the equilibrated state resemble those of the most unstable mode inferred from linear stability analysis~\cite{Green1970,Held1992,Chai2014}. An issue with this approach is that only the equilibrated state is governed by the diffusion tensor~(\ref{GMRedi}), see the derivation in~\citeA{Meunier23}. In particular, equation~(\ref{GMRedi}) indicates that the ratio of the vertical to the meridional buoyancy flux is given by the mean isopycnal slope ${\cal S}$ (adiabatic transport). As discussed in~\citeA{Eady49} and~\citeA{Vallisbook}, this constraint does not hold for an unstable eigenmode because of the non-stationary terms, the associated profiles $\overline{wb}(z)$ and $\overline{vb}(z)$ being therefore incompatible with~(\ref{GMRedi}) \cor{(in other words, one would infer a different profile for $K_{GM}(z)$ based on $\overline{vb}(z)$ or $\overline{wb}(z)$)}. We have nevertheless computed the most unstable eigenmode of the present Charney model, perturbatively for weak $\beta_*$ (see Supporting Information). As shown in Figure~\ref{fig:Charney}, the associated meridional buoyancy flux overpredicts the variations of $\overline{vb}(z)$ with depth and compares unfavorably with the present prediction~(\ref{vbexpprediction}). The most-unstable-mode approach may be better-suited for weakly nonlinear atmospheric states charaterized by a weak supercriticality $\xi=1/\beta_*$, as opposed to the present large-supercriticality oceanic situations~\cite{Jansen12}.

The success of the perturbative approximation $K_R = K_{GM}$ throughout the entire water column for case II above may come as a surprise to the reader accustomed to channel simulations, where $K_R$ typically exceeds $K_{GM}$ in the interior (see e.g. \citeA{Abernathey13}). The reason for this success is that the meridional QGPV gradient is small in the interior and around the so-called `steering levels'~\cite{Green1970,Treguier99,Smith09,Abernathey10,Abernathey13}, making the QGPV flux $\overline{vq}$ negligible there (see e.g. figure 6 of \citeA{Smith09}). One thus makes a negligible error by inferring the buoyancy and QGPV flux profiles using the approximation $K_R=K_{GM}$ throughout the entire water column.

The perturbative prediction (\ref{KGMgeneral}) for the vertical structure of the GM coefficient is simple to implement, it is easily extended to a patch of ocean subject both to zonal and meridional large-scale gradients and shear flows, it is free of adjustable parameters -- except for the overall magnitude of the transport -- and it compares very favorably with the common practice of using a depth-invariant GM coefficient. The implementation of (\ref{KGMgeneral}) in a global model should lead to a more accurate description of the stratification of the Southern Ocean, and therefore of neighboring ocean basins.
Beyond this modeling application, the physically-based vertical structure (\ref{vbgeneral}) for the buoyancy flux could be of use to infer the buoyancy flux throughout the entire water column based on near-surface data. Indeed, figure~\ref{fig:vbprofiles} shows that the prediction (\ref{vbgeneral}) allows one to propagate the value of the near-surface flux to the interior of the water column in a way that agrees closely with the full DNS profile. By contrast, propagating the near-surface information using a uniform GM coefficient would lead to the orange line in figure~\ref{fig:vbprofiles}, which at depth typically departs from the DNS profile by $40\%$ to $100\%$ depending on the situation.



\section*{Acknowledgments}
This research is supported by the European Research Council under grant agreement FLAVE 757239. The numerical study was performed using HPC resources from GENCI-CINES and TGCC (grants 2021-A0102A10803, 2022-A0122A12489 and 2023-A0142A12489).

\appendix

\nocite{Dedalus,Miquel20,Miquel21coral,Bouillaut21,miquelPRF19}
\bibliography{VSbib}

\end{document}